\documentclass{revtex4}
\usepackage{graphicx}
\usepackage{fancyhdr}
\usepackage{amsmath}
\usepackage{color}

\begin{document}

\title{B physics and Extended Higgs sectors \\ (Tauonic B decays and two Higgs doublet models)}

%

\author{Ryoutaro Watanabe}
\affiliation{Theory Group, KEK, Tsukuba, Ibaraki 305-0801, JAPAN}

\begin{abstract}
Recent experimental results on exclusive semi-tauonic $B$ meson decays, $\bar B\to D^{(*)}\tau\bar\nu$, show sizable deviations from the standard model prediction, 
while the recent result on pure-tauonic decay, $\bar B\to \tau\bar\nu$, reduces the deviation from the prediction. 
These results suggest an indirect evidence of new physics in which the structure of the relevant weak charged interaction may differ from that of the standard model. 
We study these tauonic $B$ decays in the context of extensions of the Higgs sector. 
As a result, we find that two Higgs doublet models without tree-level FCNC are unlikely to explain the present experimental data while those allowing FCNC are consistent with the data. 
 \end{abstract}

\maketitle

\thispagestyle{fancy}


\section{Introduction}
\label{Sec:Intro}
\begin{figure}[h]
\includegraphics[width=30em]{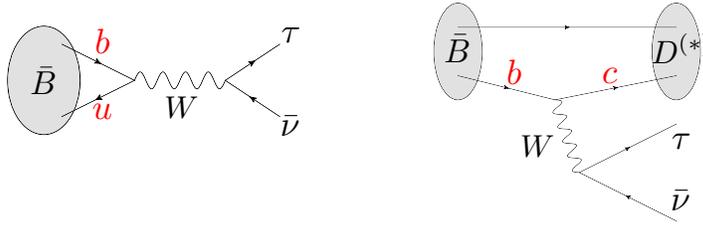}
\caption{$W$ boson contribution to the decays.}
\label{FIG:SMdiagram}
\end{figure}

Among the $B$ meson decays, $\bar B \to \tau\bar\nu$ and $\bar B \to D^{(*)} \tau\bar\nu$ contain both the heavy quark ($b$) and lepton ($\tau$) in the third generation. 
Therefore these processes are relatively sensitive to the effect of the charged Higgs bosons\cite{GH, HOU93}, 
while they are described as processes mediated by a $W$ boson in the SM as shown in Fig.~\ref{FIG:SMdiagram}. 
Then these tauonic B decays are the golden modes in the search for the charged Higgs bosons at a future super $B$ factory. 
From the experimental point of view, these decay processes are rather difficult to be identified because of two or more missing neutrinos in the final states. 
At (super) $B$ factories, however, reconstructing one of the $B$ mesons in the $e^+e^- \to \Upsilon(4S) \to B \bar B$ reaction, 
one can compare properties of the remaining particles to those expected for signal and background. 
This method allows us to identify and measure the $B$ meson decays including missing particles. 

The branching ratio $\mathcal B(\bar B \to \tau\bar\nu)$ in the SM is obtained 
by use of the inputs $|V_{ub}|=(3.38\pm0.15)\times10^{-3}$ coming from the fit of CKM triangle\cite{MyThesis} and $f_B =(191\pm9) \text{MeV}$ determined by the lattice QCD study\cite{FBconstant}. 
For a precise prediction of $\bar B \to D^{(*)} \tau\bar\nu$, it is useful to take the ratio of branching fraction to these light-leptonic decay modes. 
The ratios of the branching fractions are defined by
\begin{equation}
R(D^{(*)})\equiv\frac{\mathcal{B}(\bar B\to D^{(*)}\tau^-\bar\nu_\tau)}{\mathcal{B}(\bar B\to D^{(*)}\ell^-\bar\nu_\ell)}, 
\end{equation}
and its value in the SM is precisely evaluated by use of a heavy quark effective theory\cite{CLN}. 
The Belle and BABAR collaborations reported their new results of $\bar B \to \tau\bar\nu$ and $\bar B \to D^{(*)} \tau\bar\nu$ respectively in the last year\cite{Babar2012,BELLE2012} using the full data set. 
In Table~\ref{Tab:RD}, we summarize the experimental results and the theoretical predictions in the SM, 
where the average values are obtained by the combination of BABAR\cite{BABARPT,Babar2012} and Belle\cite{BELLE2012,BELLEST1,BELLEST2,BELLEST3} assuming the gaussian distribution.
As seen in Table~\ref{Tab:RD}, the SM is disfavored at $3.5\sigma$ in $\bar B \to D^{(*)} \tau\bar\nu$ while the result in $\bar B \to \tau\bar\nu$ is consistent with the prediction in the SM.  
Then these results imply an existence of sizable new physics effects in $\bar B \to D^{(*)} \tau\bar\nu$. 
As shown in Ref.~\cite{Babar2012}, however, one finds that these excesses cannot be explained by a charged Higgs boson in the two Higgs doublet model (2HDM) of type II at the same time. 
It is easily expected that the situation on both the results of $\bar B\to D^{(*)} \tau\bar\nu$ and $\bar B \to \tau\bar\nu$ is not suitable for the 2HDM 
because the result of $\bar B \to \tau\bar\nu$ is consistent with the prediction in the SM as explained above. 
Recently there are several studies in these decays to explain the present experimental results\cite{FKN,Sakaki,DDG,BKT,CJLP,FKNZ,OUR2012}. 
In this work, we investigate the tauonic B decays in the 2HDMs with/without flavor changing neutral current (FCNC) in the Yukawa term 
and discuss a possibility to explain the recent results of these decays within the 2HDMs. 
\begin{table}\begin{center}\begin{tabular}{c|cccc}
 \hline                                                                              &  Belle\cite{BELLE2012}  &  Average                    & Prediction (SM) \\
 \hline $\mathcal B(\bar B \to \tau\bar\nu) \times 10^4$ &  $0.72\pm0.28$               &  $1.14\pm0.23$         & $0.74\pm0.07$ \\
 \hline 
 \hline                      &  BABAR\cite{Babar2012}                               &  Average                                                          & Prediction (SM) \\
 \hline $R(D)$         &  $0.44\pm0.07$                                               &  $0.42\pm0.06$                                               &  $0.305\pm0.012$ \\
 \hline $R(D^*)$      &  \hspace{1em}$0.33\pm0.03$\hspace{1em}  &  \hspace{1em}$0.34\pm0.03$\hspace{1em}  &  \hspace{1em}$0.252\pm0.004$\hspace{1em} \\
 \hline
 \hline
\end{tabular}\end{center}
\caption{Experimental results and predictions in the SM on the measurements of $\mathcal B(\bar B \to \tau\bar\nu)$ and $R(D^{(*)})$.}\label{Tab:RD}
\end{table}

\section{Two Higgs doublet models}
As known well, the 2HDMs contribute to the tauonic $B$ meson decays and its effect is enhanced in some cases. 
In order to forbid flavor changing neutral currents (FCNC) at the tree level, a $Z_2$ symmetry is often imposed in this class of models and it results in four distinct 2HDMs\cite{THDMKanemura} . 
Their Yukawa terms are described as 
\begin{eqnarray}
 \mathcal{L}_\text{Y} &=& -\bar Q_L Y_u \tilde H_2 u_R -\bar Q_L Y_d H_2 d_R -\bar L_L Y_\ell H_2 \ell_R +\text{h.c.}\quad \text{(type I)}  \,, \\
 \mathcal{L}_\text{Y} &=& -\bar Q_L Y_u \tilde H_2 u_R -\bar Q_L Y_d H_1 d_R -\bar L_L Y_\ell H_1 \ell_R +\text{h.c.}\quad \text{(type II)}  \,, \\
 \mathcal{L}_\text{Y} &=& -\bar Q_L Y_u \tilde H_2 u_R -\bar Q_L Y_d H_2 d_R -\bar L_L Y_\ell H_1 \ell_R +\text{h.c.}\quad \text{(type X)}  \,, \\
 \mathcal{L}_\text{Y} &=& -\bar Q_L Y_u \tilde H_2 u_R -\bar Q_L Y_d H_1 d_R -\bar L_L Y_\ell H_2 \ell_R +\text{h.c.}\quad \text{(type Y)}  \,,
\end{eqnarray}
where $H_{1,2}$ are Higgs doublets defined as 
\begin{eqnarray}
 H_i = \begin{pmatrix} h_i^+ \\ (v_i +h_i^0)/\sqrt 2 \end{pmatrix}, \quad \tilde H_i = i\sigma_2 H_i, 
\end{eqnarray}
and $v_i$ denotes the vacuum expectation value (VEV) of $H_i$. 
The ratio of two VEVs is defined as $\tan \beta = v_2/v_1$ and $v=\sqrt{v_1^2+v_2^2}=246\text{GeV}$.
In type I, all masses of quarks and leptons are given by $v_2$. 
In type II, the down-type quarks and leptons acquire their masses from $v_1$, while the up-type quarks from $v_2$. 
In type X, the Higgs fields $H_2$ and $H_1$ give the masses to the quarks and the leptons respectively. 
In type Y, the masses of the down-type quarks are given by $v_1$ and other fermions obtain their masses from $v_2$.  
Under this definition $v_2$ generates up-quark masses in any types of Yukawa interaction.

\begin{table}[h]
\begin{center}\begin{tabular}{c|cccc}
 \hline 
 \hline              & Type\,I             & Type\,II            & Type\,X & Type\,Y \\
 \hline $\xi_d$ & $\cot^2 \beta$  & $\tan^2 \beta$ & $-1$       & $-1$ \\
 \hline $\xi_u$ & $-\cot^2 \beta$ & $1$                  & $1$        & $-\cot^2 \beta$ \\
 \hline 
 \hline
\end{tabular}\end{center}
\caption{Parameters $\xi_{d,u}$ in each type of 2HDMs.}\label{Tab:2HDM}
\end{table}
These 2HDMs contain a pair of physical charged Higgs bosons, which contributes to $\bar B \to \tau \bar\nu$ and $\bar B \to D^{(*)} \tau \bar\nu$ at the tree level. 
The relevant effective Lagrangian is represented as  
\begin{eqnarray}
  \mathcal{L}_\text{eff} = -2\sqrt{2}G_F V_{qb} \left( \bar{q}_L\gamma^\mu b_L\, \bar{\tau}_L\gamma_\mu (\nu_\tau)_L 
  +C_{S_1}^q \bar{q}_L b_R\,\bar{\tau}_R (\nu_\tau)_L +C_{S_2}^q \bar{q}_R b_L\,\bar{\tau}_R (\nu_\tau)_L \right),
\end{eqnarray}
with 
\begin{eqnarray}
 C_{S_1}^u = C_{S_1}^c = -\frac{m_b m_\tau}{m_{H^\pm}^2} \xi_d \,, \quad
 C_{S_2}^u = -\frac{m_u m_\tau}{m_{H^\pm}^2} \xi_u \,,\quad 
 C_{S_2}^c = - \frac{m_c m_\tau}{m_{H^\pm}^2} \xi_u \,,
\end{eqnarray}
where $q=c (u)$ corresponds to the case of $\bar B \to D^{(*)} \tau \bar\nu$ ($\bar B \to \tau \bar\nu$) and $m_{H^\pm}$ is the mass of the charged Higgs boson. 
The parameters $\xi_d$ and $\xi_u$ are presented in Table~\ref{Tab:2HDM}. 
One can see that the charged Higgs interaction corresponding to $S_1$-type affect $\bar B \to \tau \bar\nu$ and $\bar B \to D^{(*)} \tau \bar\nu$ in the same fashion. 
For the $S_2$-type operator, the contribution of the charged Higgs to $\bar B \to \tau \bar\nu$ is very suppressed due to the small up quark mass. 
As explained in Sec.~\ref{Sec:Intro}, a sizable new physics effect on $R(D^{(*)})$ is needed in order to explain the experimental results. 
Thus it is naively expected that $S_2$-type interaction in the 2HDMs is suitable to explain the recent experimental results in $\bar B \to \tau \bar\nu$ and $\bar B \to D^{(*)} \tau \bar\nu$ at the same time. 
To have a sizable charged Higgs effect, $|\xi_{d,u}|$ should be much larger than unity taking the experimental lower bound on the charged Higgs mass into account. 
Then the case of $\xi_u=1$ or $\xi_d=-1$ is not acceptable. 
The case of $\xi_u=-\cot^2\beta$ or $\xi_d=\cot^2\beta$ with $\cot^2\beta\gg1$ is unnatural since the top Yukawa interaction becomes nonperturbative.  
The requirement for the top Yukawa interaction to be perturbative results in $\tan\beta \gtrsim 0.4$ \cite{THDMKanemura}. 
Therefore, the $S_2$-type operator cannot have sizable effect on $\bar B \to D^{(*)} \tau \bar\nu$ in the 2HDM.

\begin{figure}
\includegraphics[width=33em]{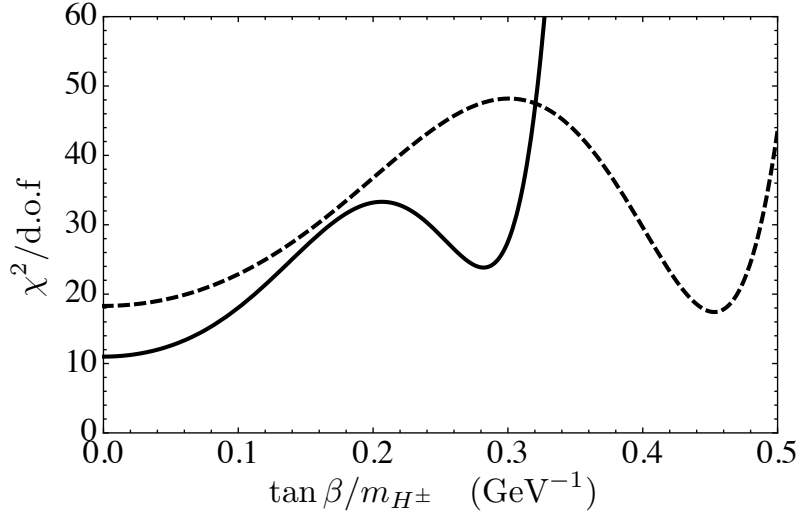}
\caption{Fit of $\tan\beta/m_{H^\pm}$ to both the results of $\mathcal B(\bar B \to \tau \bar\nu)$ and $R(D^{(*)})$ (black line), and those of only $R(D^{(*)})$ (black dashed line).}
\label{FIG:chisquare}
\end{figure}

As a consequence, the type II of $\xi_d =\tan^2\beta$ is only the case to be sizable and only $C_{S_1}^{u,c}$ in the 2HDM of type II is potentially enhanced. 
In Fig.~\ref{FIG:chisquare}, we show the $\chi^2$ fit of the charged Higgs parameter $\tan\beta/m_{H^\pm}$ to the experimental results in the 2HDM of type II. 
The black dashed line represents the $\chi^2$ fit to the results of $R(D^{(*)})$, which suggest that any value of $\tan\beta/m_{H^\pm}$ is excluded at more than $99.8\%$ confidence level (CL).  
The black solid line indicates the $\chi^2$ fit to both the results of $\mathcal B(\bar B \to \tau \bar\nu)$ and $R(D^{(*)})$, 
in which we can see that the sizable value of $\tan\beta/m_{H^\pm}$ is disfavored at more than $99.9\%$  CL.  
It is noted that the exclusion CL in the small value coming from the fit to all the tauonic B decays is smaller than that to only $R(D^{(*)})$. 
This is because the result of $\mathcal B(\bar B \to \tau \bar\nu)$ is near consistent with the SM prediction. 

\section{Tree level FCNC in Yukawa sector}
A possible solution within 2HDMs is to violate the $Z_2$ symmetry at the cost of FCNC. 
We introduce the following $Z_2$ breaking terms in the above four models: 
\begin{eqnarray}
 \Delta \mathcal{L}_\text{Y} &=& -\bar Q_L \epsilon''_u \tilde H_1 u_R -\bar Q_L \epsilon''_d H_1 d_R +\text{h.c.}\quad \text{(for type I and X)} \,, \\
 \Delta \mathcal{L}_\text{Y} &=& -\bar Q_L \epsilon''_u \tilde H_1 u_R -\bar Q_L \epsilon''_d H_2 d_R +\text{h.c.}\quad \text{(for type II and Y)} \,, \label{Eq:Z2type2}  
\end{eqnarray}
where $\epsilon''_{u,d}$ are $3\times3$ matrices that control FCNC and the quark fields are those in the weak basis. 
To obtain the charged Higgs interaction in the mass basis, first let us rotate the quark fields into 
\begin{eqnarray}
 u_{L(R)} \to U_u^{L(R)}u_{L(R)},\quad d_{L(R)} \to U_d^{L(R)} d_{L(R)}, 
\end{eqnarray}
so as to diagonalize $Y_{u,d}$. 
Then, the mass matrices are rewritten as 
\begin{eqnarray}
 M_u = \frac{1}{\sqrt2} \Big( v_2\,Y_u^D + v_1\,U_u^{L\dag}\epsilon''_uU_u^R \Big),\quad
 M_d (x,y) = \frac{1}{\sqrt2} \left( x\,Y_d^D + y\,U_d^{L\dag}\epsilon''_dU_d^R \right),
\end{eqnarray}
where $Y_{u,d}^D$ is the diagonal Yukawa matrix.  
The down-type quark mass term is represented as $M_d (v_2,v_1)$ for type I and X, and $M_d (v_1,v_2)$ for type II and Y. 
At this stage, the Yukawa terms are rewritten as 
\begin{equation}\label{Eq:FCNCIX}
 \mathcal{L}_\text{Y} +\Delta \mathcal{L}_\text{Y} = 
 -\bar u_L V_\text{CKM}^0 \left( {Y_d^D \over \sqrt2} h_2^0 +\epsilon'_u h_1^+ \right) d_R 
 +\bar d_L V_\text{CKM}^{0\dag} \left( {Y_u^D \over \sqrt2} h_2^0 +\epsilon'_u h_1^- \right) u_R +\text{h.c.\,(I,X)},
\end{equation}
\begin{equation}\label{Eq:FCNCIIY}
 \mathcal{L}_\text{Y} +\Delta \mathcal{L}_\text{Y} = 
 -\bar u_L V_\text{CKM}^0 \left( {Y_d^D \over \sqrt2} h_2^0 +\epsilon'_u h_1^+ \right) d_R 
 +\bar d_L V_\text{CKM}^{0\dag} \left( {Y_u^D \over \sqrt2} h_1^0 +\epsilon'_u h_2^- \right) u_R +\text{h.c.\,(II,Y)},
\end{equation}
where $V_\text{CKM}^0 =U_u^{L\dag}U_d^L$ is the Cabibbo-Kobayashi-Maskawa (CKM) matrix in the basis diagonalizing $Y_{u,d}$ and we define $\epsilon'_q =U_q^{L\dag}\epsilon''_qU_q^R$ (for $q=u,d$). 
In turn, rotating the quark fields into 
\begin{eqnarray}
 u_{L(R)} \to W_u^{L(R)}u_{L(R)},\quad d_{L(R)} \to W_d^{L(R)} d_{L(R)}, 
\end{eqnarray}
the mass matrices are diagonalized: 
\begin{eqnarray}
 M_u^D = \frac{1}{\sqrt2} \Big( v_2\, W_u^{L\dag}Y_u^DW_u^R + v_1\,\epsilon_u \Big),\quad 
 M_d^D (x,y) = \frac{1}{\sqrt2} \Big( x\,W_d^{L\dag}Y_d^DW_d^R + y\,\epsilon_d \Big),
\end{eqnarray}
where we define $\epsilon_q =U_q^{W\dag}\epsilon'_qW_q^R$. 
Then, rewriting the diagonal Yukawa coupling $Y_{u,d}^D$ in Eqs.~(\ref{Eq:FCNCIX}) and (\ref{Eq:FCNCIIY}) 
by use of the diagonalized quark masses $M_{u,d}^D$ and the extra coupling $\epsilon_{u,d}$ that induce FCNC, 
the physical charged Higgs and fermion interacting terms take the following form: 
\begin{eqnarray}
 \mathcal{L}_{H^\pm} =\left( \bar u_R Z_u^\dag V_\text{CKM} d_L +\bar u_L V_\text{CKM} Z_d d_R +\bar\nu_L Z_\ell \ell_R \right) H^+ +\text{h.c.} \,, 
\end{eqnarray}
where $V_\text{CKM} = W_u^{L\dag} V_\text{CKM}^0 W_d^L$ is the CKM matrix in the (true) quark mass basis. 
Table~\ref{Tab:general2HDM} shows the expressions of $Z_{u,d,\ell}$, 
where $M_{u,d,\ell}$ denote the diagonal up-type quark, down-type quark, and lepton mass matrices, 
and $\epsilon_{u,d}$ represent matrices $\epsilon''_{u,d}$ in the quark mass basis as defined above. 
\begin{table}\begin{center}\begin{tabular}{c|ccc}
 \hline 
 \hline              & $Z_u$            & $Z_d$            & $Z_\ell$  \\
 \hline Type\,I  & $\frac{\sqrt2 M_u}{v} \cot\beta -\epsilon_u \sin\beta(1+\cot^2\beta) $  & $-\frac{\sqrt2 M_d}{v} \cot\beta +\epsilon_d \sin\beta(1+\cot^2\beta) $ & $-\frac{\sqrt2 M_\ell}{v}\cot\beta$ \\
 \hline Type\,II & $\frac{\sqrt2 M_u}{v} \cot\beta -\epsilon_u \cos\beta(\tan\beta+\cot\beta) $ & $\frac{\sqrt2 M_d}{v} \tan\beta -\epsilon_d \sin\beta(\tan\beta+\cot\beta) $ & $\frac{\sqrt2 M_\ell}{v}\tan\beta$ \\
 \hline Type\,X & $\frac{\sqrt2 M_u}{v} \cot\beta -\epsilon_u \sin\beta(1+\cot^2\beta) $& $-\frac{\sqrt2 M_d}{v} \cot\beta +\epsilon_d \sin\beta(1+\cot^2\beta) $ & $\frac{\sqrt2 M_\ell}{v}\tan\beta$ \\
 \hline Type\,Y &  \hspace{0.5em}$\frac{\sqrt2 M_u}{v} \cot\beta -\epsilon_u \cos\beta(\tan\beta+\cot\beta) $\hspace{0.5em} & \hspace{0.5em}$\frac{\sqrt2 M_d}{v} \tan\beta -\epsilon_d \sin\beta(\tan\beta+\cot\beta) $\hspace{0.5em} &\hspace{0.5em}$-\frac{\sqrt2 M_\ell}{v}\cot\beta$\hspace{0.5em} \\
 \hline
 \hline
\end{tabular}\end{center}
\caption{The matrices $Z_{u,d,\ell}$ in each type of the 2HDM in the presence of the $Z_2$ breaking terms.}\label{Tab:general2HDM}
\end{table}

The FCNC in the down-quark sector is strongly constrained from several B meson decay processes, so that $\epsilon_d$ is negligible. 
On the other hand, constraints on the FCNC in the up quark sector are rather weak. 
Recently the 2HDM of type II allowing FCNC, which is called as type III in the standard convention, in the up quark sector is studied to explain $\bar B \to \tau \bar\nu$ and $\bar B \to D^{(*)} \tau \bar\nu$ at the same time\cite{CGK}. 
As can be seen in Table~\ref{Tab:general2HDM}, we find that the $S_2$ operator in not only type II but also type X might be significant for large $\tan\beta$. 
The corresponding Wilson coefficient is given by 
\begin{eqnarray}
 C_{S_2}^c \simeq \frac{V_{tb}}{\sqrt2 V_{cb}} \frac{v m_\tau}{m_{H^\pm}^2} (\epsilon_u^{tc})^* \sin\beta\tan\beta, \quad
 C_{S_2}^u \simeq \frac{V_{tb}}{\sqrt2 V_{ub}} \frac{v m_\tau}{m_{H^\pm}^2} (\epsilon_u^{tu})^* \sin\beta\tan\beta. 
\end{eqnarray} 
In this case, the different components $\epsilon_u^{tu}$ and $\epsilon_u^{tc}$ of the FCNC matrix are involved in $\bar B \to \tau \bar\nu$ and $\bar B \to D^{(*)} \tau \bar\nu$. 
As seen in Ref.~\cite{OUR2012}, the current experimental results of $\bar B \to D^{(*)} \tau \bar\nu$ are described by the 2HDM of type II or X with FCNC provided that $|\epsilon_u^{tc}| \sim 1$, 
while $\epsilon_u^{tu}$ is highly suppressed because of the enhancement factor $V_{tb}/V_{ub}$. 
If this is the case, we expect sizable deviations in polarizations $P_\tau(D^{(*)})$ and $P_{D^*}$ from the SM as suggested in Ref.~\cite{OUR2012,TANAKA,OUR}.

The large component $\epsilon_u^{tc}$ affects the top quark decay to charm quark. 
The Lagrangian which induce the FCNC process in the up-quark sector is given by  
\begin{equation}
 \mathcal{L}_{h,H,A} = \bar u_L \left[ \frac{\cos(\alpha-\beta)}{\sqrt2\sin\beta}\,h +\frac{\sin(\alpha-\beta)}{\sqrt2\sin\beta}\,H +\frac{i}{\sqrt2\sin\beta}\,A\right] \epsilon_u u_R +\text{h.c.} \,, 
\end{equation}
where $h(H)$ is the light(heavy) physical CP even Higgs, $A$ is the CP odd Higgs, and $\alpha$ is a mixing angle between these two physical CP even Higgs. 
For example, let us consider the decay process $t \to c h$. 
The decay rate $\Gamma(t\to ch)$ is calculated as 
\begin{equation}
 \Gamma(t\to ch) = \frac{|\epsilon_u^{tc}|^2}{64\pi} \frac{\cos^2(\alpha-\beta)}{\sin^2\beta} m_t \left(1-\frac{m_h^2}{m_t^2} \right)^2\,, 
\end{equation}
where $m_h \simeq126\text{GeV}$ is obtained by the ATLAS\cite{ATLAS} and CMS\cite{CMS} experiments. 
The ratio of $\Gamma(t\to ch)$ and $\Gamma(t\to bW)$, which is the dominant decay process, is naively evaluated as 
\begin{equation}
 {\Gamma(t\to ch) \over \Gamma(t\to bW)} \simeq 0.12 \frac{|\epsilon_u^{tc}|^2\cos^2(\alpha-\beta)}{\sin^2\beta} \,.  
\end{equation}
Therefore, the large value $|\epsilon_u^{tc}| \sim 1$ desired for $\bar B \to D^{(*)} \tau \bar\nu$ gives at most around $10\%$ of the branching ratio. 
The top quark decay to charm quark might be difficult to measure at the LHC experiment due to charm identification and good target for the ILC experiment.

\section{Summary and Discussion}
The latest results of $R( D^{(*)} )$ show sizable deviations from the SM prediction, while that of $\mathcal B (\bar B\to \tau\bar\nu)$ turns out to be consistent with the SM prediction with some accuracy. 
In response, we have studied the effects on $\bar B \to \tau \bar\nu$ and $\bar B \to D^{(*)} \tau \bar\nu$ in the 2HDMs. 
We have shown that the four distinct 2HDMs on which the $Z_2$ symmetry is imposed in the Yukawa term 
cannot have large contributions to $\bar B \to D^{(*)} \tau \bar\nu$ except for the type II in the operator $\bar{c}_L b_R\,\bar{\tau}_R (\nu_\tau)_L$. 
The 2HDM of type II, however, is unlikely to explain the experimental results of $\bar B \to \tau \bar\nu$ and $\bar B \to D^{(*)} \tau \bar\nu$ as we have represented in Fig.~\ref{FIG:chisquare}. 

In order to solve this situation within 2HDMs, we have studied the 2HDMs allowing the FCNC in the Yukawa term. 
As a result, we have shown that the 2HDMs of type II and X, (i.e. type III) can explain the experimental results at the cost of the Higgs induced FCNC in the top quark decays. 
In addition, we have estimated the contribution of the Higgs induced FCNC to the decay $t \to ch$ when the sizable FCNC effect, which is needed for $\bar B \to D^{(*)} \tau \bar\nu$, exists. 
In a future experiment such as the ILC, this process might be measured and useful to confirm the scenario discussed in this work.

\begin{acknowledgments}
I am very grateful to Minoru Tanaka for his great advises, encouragements, stimulating discussions and collaborations. 
I would like to thank Andrey Tayduganov for useful comments and discussions. 
This work is supported in part by the Grant-in-Aid for Science Research, Ministry of Education, Culture, Sports, Science and Technology, Japan, under Grant No.~248920. 
\end{acknowledgments}

\end{document}